\documentclass[11pt,a4paper]{article}

\usepackage{amsmath}
\usepackage{lastpage,fancyhdr,graphicx}\graphicspath{{./FIGS/}}
\usepackage{hyperref}
\usepackage{amssymb}
\usepackage{graphics}
\usepackage{xcolor}
\usepackage[normalem]{ulem}
\usepackage{pdfpages}
\usepackage{epstopdf}
\usepackage{authblk}
\newcommand{\eps}{\varepsilon}
\title{Neuronal mechanisms for sequential activation of memory items: dynamics and reliability}
\author[2,3]{Elif K{\"o}ksal-Ers{\"o}z}
\author[1,5]{Carlos Aguilar}
\author[2, 4]{Pascal Chossat}
\author[2,4]{Martin Krupa}
\author[1]{Fr\'ed\'eric Lavigne}
\affil[1]{\small Universit\'e C\^ote d'Azur, CNRS-BCL,  Nice, France} 
\affil[2]{\small Project Team  MathNeuro, INRIA-CNRS-UNS,  Sophia Antipolis, France}
\affil[3]{\small Signal and Image Processing Laboratory,  INSERM 1099, University of Rennes, Rennes, France }
\affil[4]{\small Universit\'e C\^ote d'Azur,  Laboratoire Jean-Alexandre Dieudonn\' e, 
 Nice, France}
\affil[5]{\small Amaris, 950 Route des Colles,  Biot, France}

\begin{document}

\maketitle

\begin{abstract}
\noindent In this article we present a biologically inspired model of activation of memory items in a sequence. Our model produces 
two types of sequences, corresponding to two different types of cerebral functions: activation of regular or irregular sequences.
The switch between the two types of activation occurs through the modulation of biological parameters, without altering the connectivity matrix. Some of the parameters included in our model are neuronal gain, strength of inhibition, synaptic depression and noise.
We investigate how these parameters enable the existence of sequences and influence the type of sequences observed. 
In particular we show that synaptic depression and noise drive the transitions from one memory item to the next and neuronal gain controls the switching between regular and irregular (random) activations.  
\end{abstract}

\section{Introduction}

The processing of sequences of items in memory is a fundamental issue for the brain to generate sequences of stimuli necessary for goal-directed behavior \cite{Pezzulo2014}, language processing \cite{Burgess1999} \cite{Lavigne2011}, musical performance \cite{Rohrmeier2012} \cite{Zatorre2007}, thinking and decision making \cite{Graziano2011} and more generally  prediction \cite{Bubic2010}, \cite{Meyer2011}, \cite{Kok2012}. Those processes rely on priming mechanisms in which a triggering stimulus (e.g. a prime word) activates items in memory corresponding to stimuli not actually presented (e.g. target words; \cite{Brunel2009}, \cite{Lerner2012}. A given triggering stimulus can generate two types of sequences: on the one hand, the systematic activation of a same sequence is required to repeat reliable behaviors \cite{Velizcuba2015}, \cite{Conway2001}, \cite{Buhusi2005}, \cite{Eagleman2012}, \cite{Xu2012}; on the other hand, the generation of variable sequences is necessary for the creation of new behaviors \cite{Buckner2008}, \cite{Christoff2009}, \cite{Guilford1950}, \cite {Abraham2012}, \cite{Gonen2013}. Hence the brain has to face two opposite constraints of generating repetitive sequences or of generating new sequences. Satisfying both constraints challenges the link between the types of sequence generated by the brain and the relevant biological parameters. Can a neural network with a fixed synaptic matrix switch behavior between reproducing a sequence and produce new sequences? And which neuronal mechanisms are sufficient for such switch in the type of sequence generated? The question addressed here is how changes in neuronal noise, short-term synaptic depression and neuronal gain make possible either repetitive or variable sequences.

Neural correlates of sequence processing involve cerebral cortical areas from V1 \cite{Xu2012} \cite{Gavornik2014} and V4 \cite{Eagleman2012} to prefrontal, associative, and motor areas \cite{Jenkins1994} \cite{Sakai1998}. The neuronal mechanisms involve a distributed coding of information about items across a pattern of activity of neurons \cite{Hung2005} \cite{Young1992} \cite{Kreiman2006} \cite{QuianQuiroga2010} \cite{QuianQuiroga2016}. In priming studies, neuronal activity recorded after presentation of a prime image shifts from neurons active for that image to neurons active for another image not presented, hence beginning a sequence of neuronal patterns \cite{Miyashita1988} \cite{Erickson1999} \cite{Rainer1999} \cite{Reddy2015}. Those experiments report that a condition for the shift between neuronal patterns of activity is that stimuli have been previously learned as being associated. Considering that the synaptic matrix codes the relation between items in memory \cite{Yakovlev1998} \cite{Weinberger1998}, computational models of priming have shown that the activation of sequences of two populations of neurons rely on the efficacy of the synapses between neurons from these two populations \cite{Brunel1996} \cite{Lavigne2001} \cite{Lavigne2002} \cite{Mongillo2003} \cite{Brunel2009}. 

Turning to longer sequences, many of the models studied to date rely on the existence of steady patterns (equilibria) of saddle type, which allow for transitions from one memory item to the next \cite{Bick2012} \cite{Katkov2017} \cite{Aguilaretal2017}. Such models are well suited for reproducing systematically a same unidirectional sequence: as time evolves neuronal patterns are activated in a systematic order. These works show that the generation of directional sequences relies on the asymmetry of the relations between the populations of neurons that are activated successively. Regarding the order of populations n, n+1, n+2 in a sequence, the directionality of the sequence is obtained thanks to two properties of the synaptic matrix. First, the synaptic efficacy increases with the order of the populations, that is efficacy is weaker between populations one and two than between populations two and three \cite{Aguilaretal2017} \cite{Velizcuba2015}. Second, the amount of overlap increases with the order of populations \cite{Katkov2017}. Indeed, individual neurons respond to several different stimuli \cite{Rolls1995} \cite{Tamura2001} \cite{Tsao2006} and two populations of neurons coding for two items can share some active neurons \cite{Fujimichi2010} \cite{QuianQuiroga2012}. Models have proposed a Hebbian learning mechanism that determines synaptic efficacy as a function of the overlap between the populations \cite{Tsodyks1990}. In models the amount of overlap codes for the association between the populations and determines their order of activation in a sequence \cite{Lerner2012} \cite{Lerner2014} \cite{Katkov2017} \cite{Aguilaretal2017}. These works identify sufficient properties of the synaptic matrix to generate systematic sequences. However such properties of the synaptic matrix may not be necessary and neuronal mechanisms may also be sufficient to generate sequences. 

In this work we consider the case of fixed synaptic efficacy and fixed overlap to focus on sufficient neuronal mechanisms that underlie the type of sequence. The present study mathematically analyses a new and more general type of sequences in which the states of the network do not need to reach saddle points. The model is based on a more general mechanism of transition from one memory item to the next, with the saddle pattern replaced by a saddle-sink pair (see Ashwin and Postelthwaite \cite{ashwin-postlethwaite}, for a prototype of this mechanism of transition). As time evolves the sink and saddle patterns become increasingly similar,
so that even a small random perturbation can push the system past the saddle to the next memory item. In the model those new dynamics alleviate constraints on the synaptic matrix by allowing sequences that form spontaneously with the transitions obtained between populations related through fixed overlap, without theoretical or practical restriction on the length of the sequences. We investigate how changes in  parameters with a clear biological meaning such as neuronal noise, short-term synaptic depression and neuronal gain can control the reliability of the sequences. 

\section{Results}
\subsection{The model\label{sec:model}}
The focus of this paper is to present a mechanism of sequential activation of memory items in the absence of either increasing overlap,
or increasing synaptic conductance, or any other feature forcing directionality of the sequences.
We present this mechanism in the context of a simple system, however the idea is general and can be implemented in detailed models.
We use the neural network model of the form
\begin{align}
\label{eq:xi'}
\dot{x_i} = x_i(1-x_i)\left(-\mu x_i -\lambda \sum_{j=1}^N x_j+ \sum_{j=1}^N J_{i,j}^{max} s_j x_j \right)+\eta\\\label{eq:si'}
\dot{s_i}=\frac{1-s_i}{\tau_r}-Ux_is_i~~~~(i=1,\dots,N),
\end{align} 
as in \cite{Aguilaretal2017},
with the variables $x_i\in [0, 1]$ representing normalized averaged firing rates of excitatory neuronal populations (units), and $s_i\in [0, 1]$  controlling short term synaptic depression (STD). 
The limiting firing rates $x_i=0$ and $x_i=1$ correspond respectively to the rest and excited states of unit $i$. Any set $(x_1,\dots,x_N)$ with $x_i=0$ or $1$ ($i=1,\dots,N$) defines a steady, or equilibrium, {\em pattern} for the network. In the classical paradigm the learning process results in the formation of stable patterns of the network. Retrieving memory occurs when a cue puts the network in a state which belongs to the basin of attraction of the learned pattern.\\ 
Eq. \eqref{eq:xi'} is usually formulated using the activity variable $u_i$ (average membrane potential) rather than $x_i$, and $x_i$ is related to $u_i$ through a sigmoid transfer function. Our formulation in which the inverse of the  sigmoid is replaced by a linear function with slope $\mu$, was shown to be convenient for finding sequential retrievals of learned patterns, see \cite{Aguilaretal2017}.\\
The parameters in \eqref{eq:xi'} are $\mu$ (or its inverse $\gamma=\mu^{-1}$ which is the gain, supposed identical, of the units, or slope of the activation function of the neuron \cite{SalinasThier2000}), $\lambda$ the strength of a non-selective inhibition (inhibitory feedback due to excitation of interneurons) and $J_{i,j}^{max}$ the maximum weight of the connexion from unit $j$ to unit $i$. Finally, $\eta$ is a noise term which can be thought of as a flucutation of the firing rate due to random presence or suppression of spikes. In our simulations we considered white noise with the additional constraint of pointing towards the interior of the interval $[0,1]$. Other types of noise can be chosen, this does not affect the mechanisms which we have invertigated.\\
Short-term depression reported in cortical synapses \cite{Tsodyks1997} rapidly decreases the efficacy of synapses that transmit the activity of the pre-synaptic neuron. This is modeled by eq. \eqref{eq:si'} where $\tau_r$ is the time constant of the synapse and $U$ is the fraction of used synaptic resources. For an active unit with initially maximal synaptic strength $s_i=1$, $s_i$ decays towards the value $S=(1+\rho)^{-1}$ where $\rho=\tau_rU$. The parameter $\rho$ characterizes the STD.\\  
The main difference in the model between this paper and \cite{Aguilaretal2017} is the form of the matrix of excitatory connections $J^{\rm max}$:
\begin{align}
\label{Jmax}
J^{max}=
\begin{bmatrix} 
1& 1& 0&\dots&0\\
1& 2& 1 &\ddots& \vdots \\
0& \ddots& \ddots&\ddots&0\\
\vdots& \ddots &1&2&1\\
0& \dots &0&1&1\\
\end{bmatrix}_{N \times N}.
\end{align}
This matrix is derived by the application of the simplified learning rule of \cite{Tsodyks1998} (details provided in  
\cite{Aguilaretal2017}) using the collection of learned patterns
\begin{align}
\label{eq:xi^p}&\xi^i=(0,\dots,0,1,1,0,\dots,0)~,~i=1,\dots,P
\end{align}
where the two excited units are $i$ and $i+1$. Conditions for the stability of these patterns in the absence of STD were derived in \cite{Aguilaretal2017}.
Note that the overlap between $\xi^i$ and $\xi^{i+1}$ is constant (one unit).
By the application of the learning rule the coefficients of $J^{max}$ are given by the formula:
\begin{align}
\label{eq:Jmaxrule}
J_{i,j}^{max}  = \sum_{k=1}^P \xi_i^k\xi_j^k. 
\end{align}
Consequently the matrix $J^{max}$ is made up of identical (1 2 1) blocks along the diagonal, so that there is no increase in either overlap or the synaptic efficacy (weight) along any possible chain.  We prove mathematically and verify by numerics  that \eqref{eq:xi'} admits a chain of latching dynamics passing through the patterns $\xi^i$, $i=1,\ldots n-2$, either in forwards or in backwards direction depending on the activation, as well as shorter chains. The simplest way to switch dynamically from the learned pattern $\xi^i$ to $\xi^{i+1}$ is by having a mechanism such that unit $i$ passes from excited to rest state, then unit $i+2$ passes from rest to excited state. STD can clearly result in the inhibition of unit $i$. However in the framework of \cite{Aguilaretal2017} it was not possible to obtain the spontaneous excitation of unit $i+2$ with the connectivity matrix \eqref{Jmax}, because it was required that the upper and lower diagonal coefficients of $J^{max} $ be strictly increasing with the order $i$. 
Connectionist models have shown the effects of fast synaptic depression on semantic memory \cite{Huber2003} and on priming \cite{Lerner2012} \cite{Lerner2014}. Fast synaptic depression contributes to deactivation of neurons initially active in a pattern -- because they activate less and less each other -- in favor of the activation of neurons active in a different but overlapping pattern -- because newly activated neurons can strongly activate their associates in a new pattern. The combination of neuronal noise and fast synaptic depression enables latching dynamics in any direction depending on the initial bias due to random noise. Indeed, when the parameters lie within a suitable range, the action of STD has the effect of creating a "dynamic equilibrium" with a small basin of attraction. This dynamic equilibrium could be $\xi^i$, $\hat\xi^i$ (the pattern in which only unit $i+1$ is excited) or an intermediate pattern for which the value of $x_i$ is between $0$ and $1$. Subsequently the noise allows the system to eventually jump to $\xi^{i+1}$, the process being repeated sequentially between all or part of the learned patterns.  This noise-driven transition is what we call an {\em excitable connection} by reference to a similar phenomenon discussed in \cite{ashwin-postlethwaite}. Chains of excitable connections can also be activated or terminated by noise. 
Last but not least we show that our system, depending on the neuronal gain $\gamma$, will follow the sequence indicated by the overlap or execute a random sequence of activations. Changes in neuronal gain change the sensitivity of a neuron to its incoming activation (\cite{SalinasThier2000}, \cite{SalinasSejnowski2001}, \cite{Silver2010}), and are reported to impact contextual processing \cite{Aston-Jones2005} to enhance the quality of neuronal representations \cite{Servan-Schreiber1990} and to modulate activation between populations of neurons to reproduce priming experiments (\cite{Lavigne2008}). Here we show how changes in neuronal gain switches the networks behavior between repetitive (reliable) sequences and variable (new) sequences. 

We proceed to present the results in more detail, as follows.
In Section \ref{sec-N8} we present simulations for the network with $N=8$, which serves as an example of the more general construction. 
In Section \ref{sec-dyn} we sketch the methods we use to search for or verify the existence of the chains. In Section \ref{sec-more} we discuss irregular chains of random activations versus regular chains defined by the overlap. 

\subsection{Case study: a system with $N=8$ excitable units}\label{sec-N8}

We consider sequences of seven learned patterns $\xi^{1},...,\xi^{7}$ (named ABCDEFG) encoded by eight units $x_{1},..., x_{8}$. 
The sequence represents the sequential activation of pairs of units 1-2, 2-3, 3-4, 4-5, 5-6, 6-7 and 7-8, corresponding to patterns A and B, B and C, etc. with an overlap of one unit between them. Learning is reported to rely on changes in the efficacy of the synapses between neurons \cite{Kandel2001} through long term potentiation (LTP) and long term depression (LTD) \cite{Alberini2009} \cite{Takeuchi2014} \cite{Nabavi2014}. As a consequence, LTP/LTD potentiates/depresses synapses between units coding for patterns as a function of their overlap, that is synapses between units coding for overlapping patterns are more potentiated. Due the constant overlap, all synapses between overlapping patterns are equal. Note that the matrix $J^{max}$ \eqref{Jmax} is learned as a function of the overlap between patterns without imposing any sequences. A consequence is that learning of independent pairs of patterns generates a matrix that allows for the activation of sequences.\\
A system of $N=8$ excitatory units can encode $P=N-1=7$ regular patterns in $J^{max}$ \eqref{Jmax}. Encoded memory items can be retrieved either spontaneously (noisy environment) or when the memory network is triggered by an external cue \cite{Katkov2017} \cite{Romanietal2013}. Units $x_1$ and $x_8$ are the least self-excited units with $J^{max}_{1,1} =J^{max}_{8,8}=1$, thus it is very unlikely to active them unless they are part of the initial activity state. Hence, the longest chain has $P-1=6$ consecutive patterns.
\subsubsection{Directional sequences from a stimulus-driven pattern in the sequence}\label{sec:stimulusdriven}
\begin{figure}[h!]
\center
\includegraphics[scale=0.8]{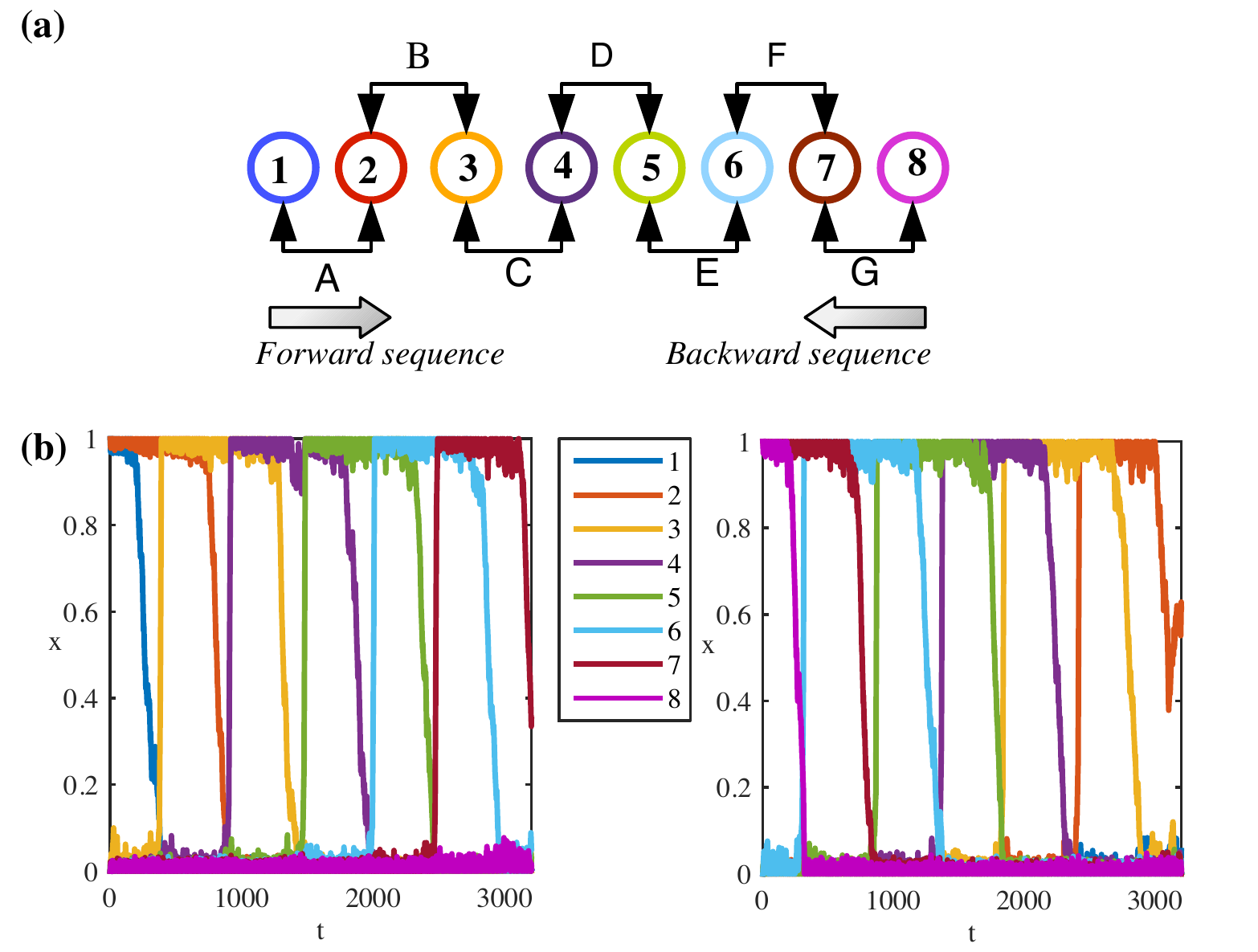}
\caption{Directional sequences of an endpoint stimulus-driven system. \textbf{(a)} Each numbered circle represents a neuron. Consecutive units encode a pattern. Except for $x_1$ and $x_8$, each unit participates in two patterns. A forward sequence is the activation of units in increasing order. A backward sequence is the activation of units in decreasing order. \textbf{(b)} Left panel: System initialised from the pattern A follows the forward sequence until the pattern F. Right panel: System initialised from the pattern G follows the backward sequence until the pattern B. Same colour code is used to represent units' indices in \textbf{(a)} and \textbf{(b)}. Parameters: $\mu = 0.40714, \lambda = 0.50714, I=0, \tau_r=900, U=0.002, \eta = 0.02$.}
\label{fig:firstLastPattern}
\end{figure}

\begin{figure}[h!]
\center
\includegraphics[scale=0.8]{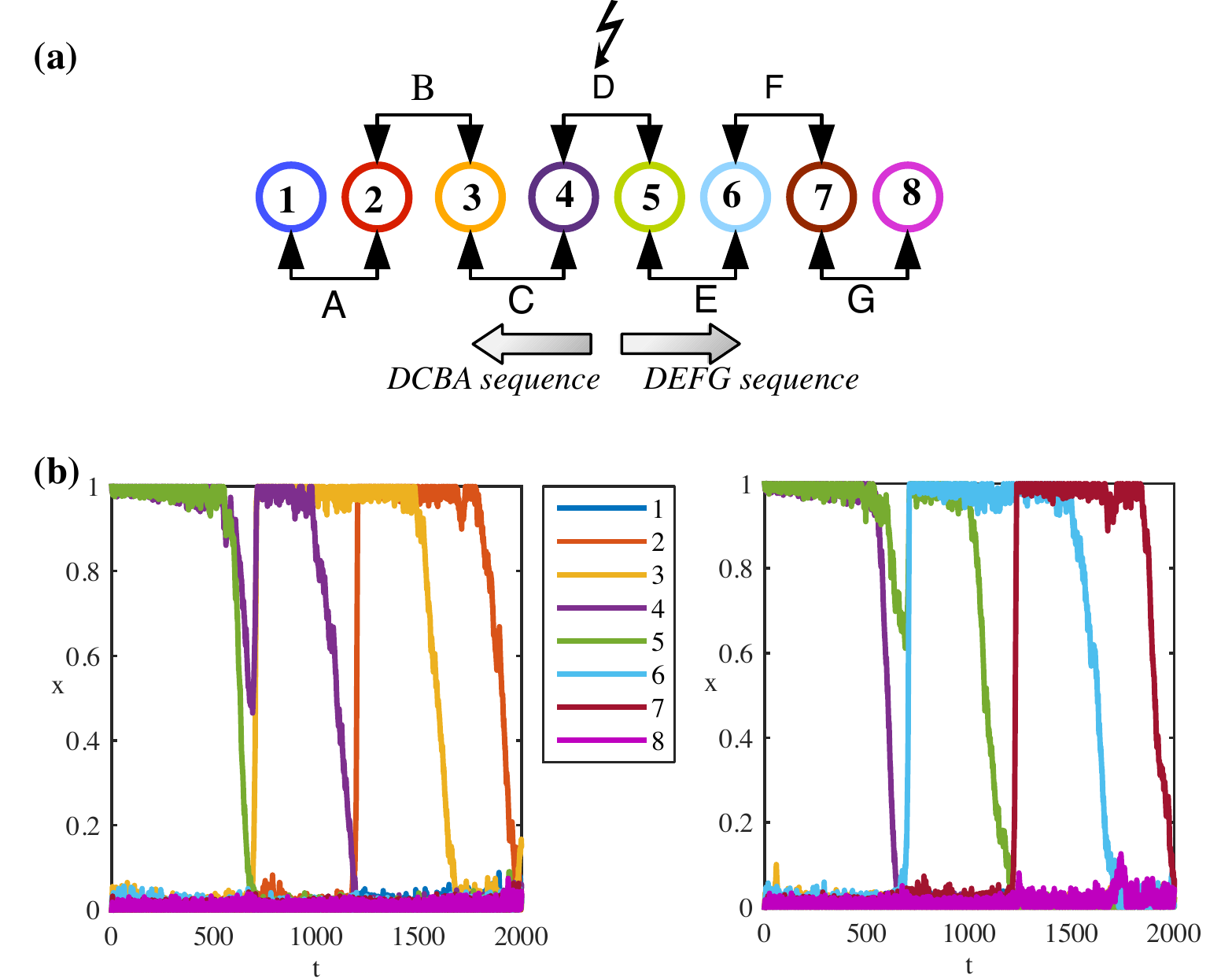}
\caption{Directional sequences of a midpoint stimulus-driven system. \textbf{(a)} Each numbered circle represents a unit. Consecutive neurons encode a pattern. Except for $x_1$ and $x_8$, each unit participates in the encoding for two patterns. When the system initialised from the pattern D, it follows either the ``DCBA" or ``DEFG" sequence. \textbf{(b)} Left panel: System initialised from the pattern D follows the ``DCBA" sequence until the pattern B. Right panel: System initialised from the pattern D follows ``DEFG" sequence until the pattern F. The same colour code is used to represent units' indices in \textbf{(a)} and \textbf{(b)}. Parameters: $\mu = 0.40714, \lambda = 0.50714, I=0, \tau_r=900, U=0.002, \eta = 0.02$.} 
\label{fig:middlePattern}
\end{figure}

Starting from the first pattern A, the directional activation corresponds to the sequence ABCDEFG (Fig. \ref{fig:firstLastPattern}-a left panel). The forward direction is imposed by $J^{max}$ because $x_1$ is less excited since $J^{max}_{1,1}=1$. Hence, while the synaptic variables $s_1$ and $s_2$ are equal and decreasing together as the system lies in the vicinity of $\xi^1$, $x_1$ is deactivated before $x_2$. In the same interval of time $s_2<s_3$ and $s_2-s_3$ increases so that $x_2$ becomes unstable before $x_3$ and the system now may converge to $\xi^2$. The process can be repated between $\xi^2$ and $\xi^3$ etc. Similarly, starting from the last pattern G gives the reverse direction (GFEDCBA) to the system (Fig. \ref{fig:firstLastPattern}-b right panel).\\ 
Initialising the system from a middle pattern $\xi^i$ doesn't introduce any direction, since the two active units of $\xi^i$ are equally excited. While their synaptic variables are decreasing together, depending on the noise at the moment when $\xi^i$ becomes unstable, either $\xi^{i-1}$ or $\xi^{i+1}$ is activated with equal probabilities. Figure \ref{fig:middlePattern} shows the response of the system starting from a mid-point pattern D. The activated sequence can go in either direction DEFG or its reverse DCBA. The \textit{random} choice for a sequence is driven by a bias in the noise at the time of stimulus-driven activation of the mid-point pattern.
%

%
%
\subsubsection{\label{sec:noisedriven} Noise-driven random sequence from a mid-point pattern in the sequence}
\begin{figure}[h!]
\center
\includegraphics[scale=0.8]{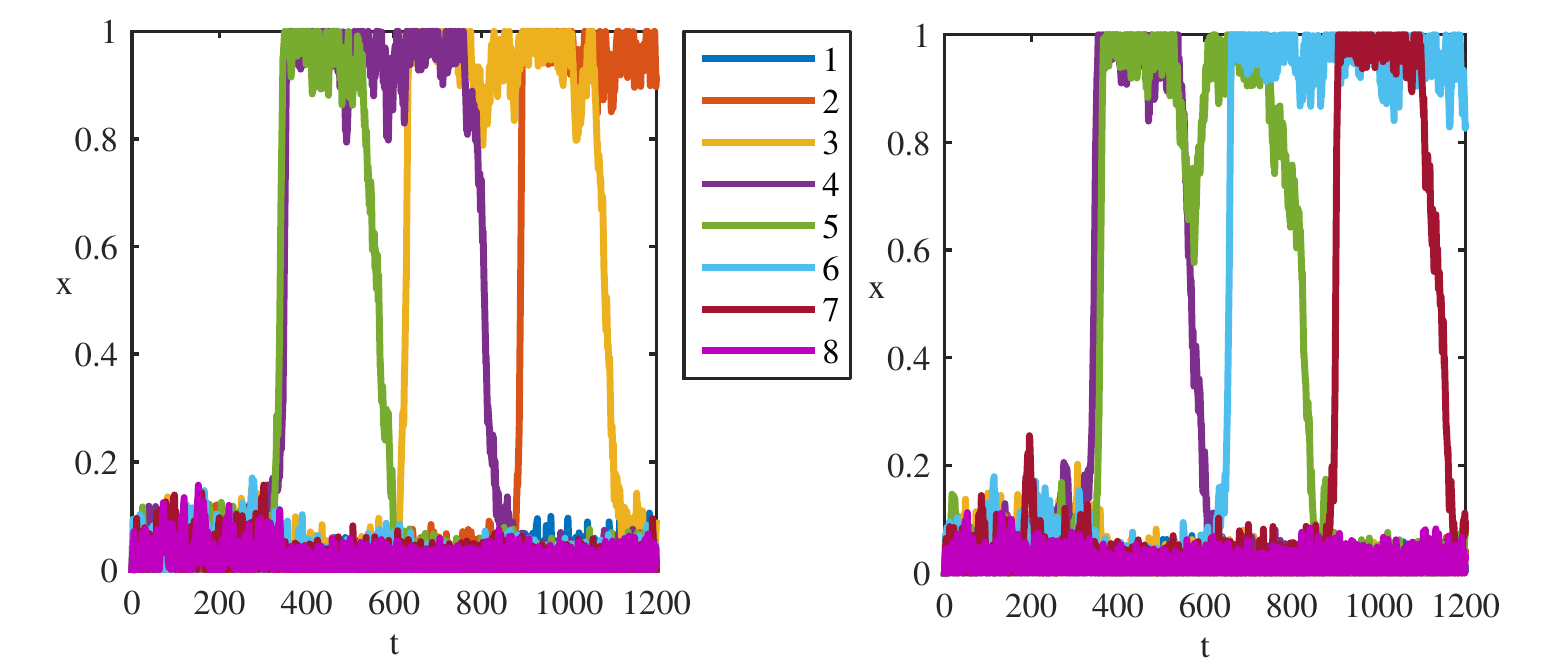}
\caption{System activated spontaneous by random noise can move in backward (left panel) or forward (right panel) directions. Parameters: $\mu = 0.20714, \lambda = 0.50714, I=0, \tau_r=300, U=0.006, \eta = 0.04$.} 
\label{fig:noiseDrivenPattern}
\end{figure}

The units that participate in two patterns (overlapping units) have stronger self-excitation as it is manifested by the diagonals of $J^{max}$. These units ($x_i, i \neq \{1,8\}$) are likely to be excited by random noise and they can activate others that they encode a pattern with. After a pattern $\xi^i$ or the associated intermediate pattern $\hat\xi^i = (0,\dots,0,1,0,\dots,0)$ being randomly excited by noise, the system can follow either $\xi^{i-1}$ or $\xi^{i+1}$. The robustness of activity depends on the system parameters. Figure \ref{fig:noiseDrivenPattern} shows an example of spontaneous activation of a mid-point pattern D where the directional oriented sequence can be either DEFG or its reverse DCBA. Similar to the system initialised from a middle pattern, the \textit{random} choice for a direction is driven by a bias in the noise at the time of noise-driven activation.

\subsubsection{Sensitivity of the dynamics upon parameter values}\label{sec:simulations}
\begin{figure}[h!]
\centering
\includegraphics[width=0.9\linewidth]{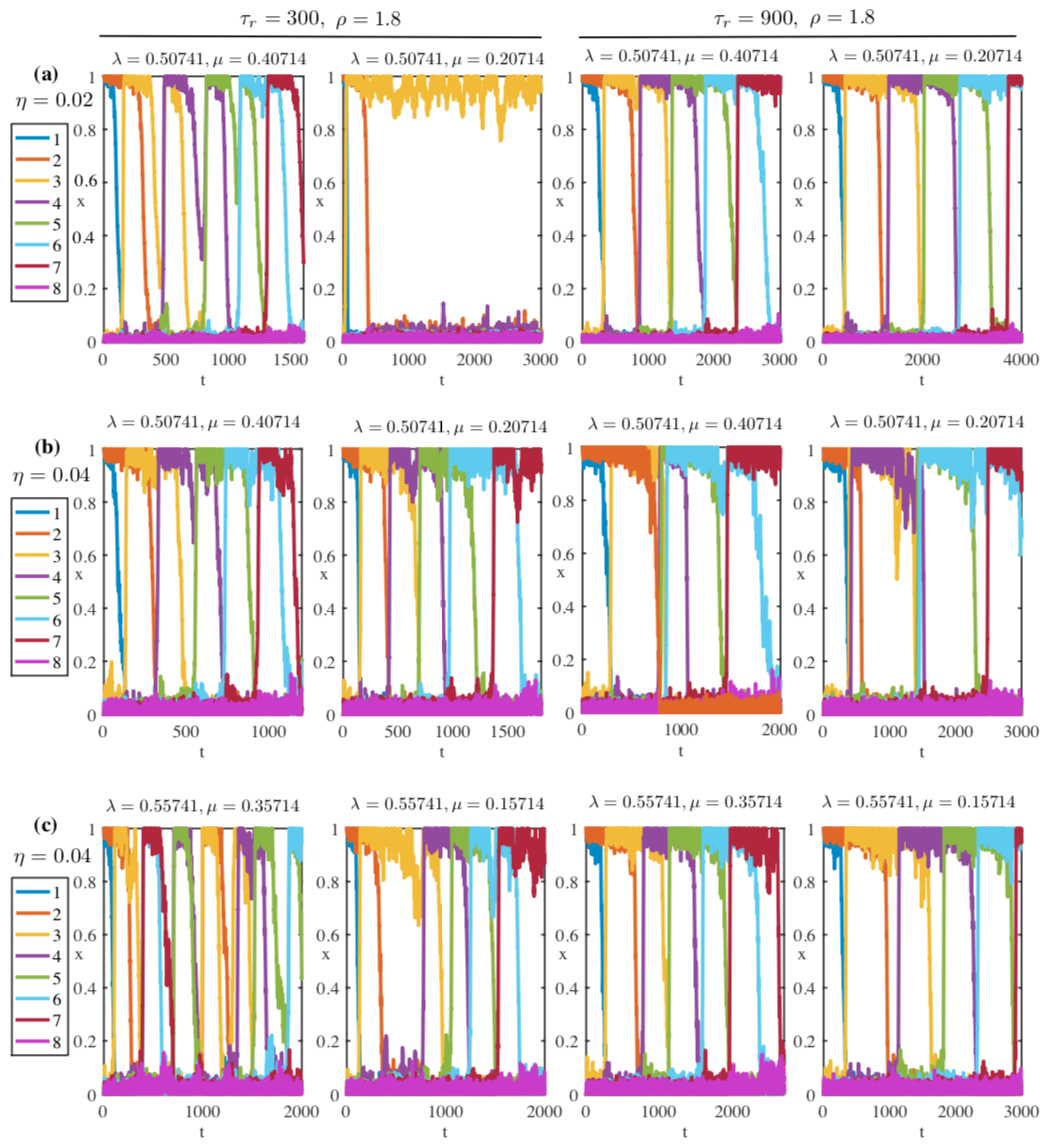}
\caption{Response of the system initialised from pattern A to different levels of noise. Synaptic variables are faster along the first two columns ($\tau_r = 300$) than the last two columns ($\tau_r = 900$).  \textbf{Row-(a)} The system with fast synapses and weak perturbation ($\eta = 0.02$) can follow the longest sequence from A to F for $\mu = 0.40714$ but not for $\mu = 0.20714$, while the slow synapses can trigger the longest sequence in either case. Increasing the noise amplitude to $\eta = 0.04$ (\textbf{row-(b)}) enables the activation of the whole sequence but slow synapses give very short patterns or 3 co-active units. While increasing the inhibition (\textbf{row-(c)}) regulates the transition for slow synapses, the system with high values of $\mu$ and fast synapses randomly activates learned patterns and yields short sequence. On the other hand, the system with small values of $\mu$ and fast synapses can preserve a regular sequence.}
\label{fig:parameterEffects}
\end{figure}

We have seen that patterns can be retrieved sequentially when the system is triggered by a cue or spontaneously by noise. However the effectiveness of this process depends on the values of the parameters in equations \eqref{eq:xi'}-\eqref{eq:si'}. The dynamics of the system can follow part of the sequence, then either terminate on one pattern $\xi^i$ with $i<N-1$, or converge to a non learned pattern.  
Moreover we identified two different dynamical scenarios by which a sequence can be followed, depending mainly on the value of $\mu$ (or gain $\gamma=\mu^{-1}$). This will be analyzed in Section \ref{sec-dyn}. Here we comment on numerical simulations which highlight the dependency of the sequences upon parameter values.\\
Figures \ref{fig:parameterEffects}abc show time series of the full or partial completion of sequences of retrievals (for $N=8$ units) for two different values of noise amplitude $\eta=0.02$ (first row) and $\eta=0.04$ (second and third row). In each case the two first columns show statistics with STD parameters $\tau_r=300$ and $U=0.006$ while the two last columns correspond to the choice $\tau_r=900$ and $U=0.002$. Recall that $\tau_r$ gives the speed at which synaptic variables decay. By taking the product $\rho=\tau_rU$ constant we ensure that the synaptic variables $s_i$ decay to the same value $S$ in both cases (see \ref{sec:model}). In \ref{fig:parameterEffects}a, \ref{fig:parameterEffects}b the global inhibition coefficient $\lambda$ is set at $0.50741$ in row two and $\lambda=0.55741$ in row three. For each choice of STD parameters the value of $\mu$ differs between the left and the right columns: $\mu=0.40714$ on the left column and $\mu=0.20714$ on the right column for Figs \ref{fig:parameterEffects}a, \ref{fig:parameterEffects}b, $\mu=0.35714$ on the left column and $\mu=0.15714$ on the right column for Fig \ref{fig:parameterEffects}c.
Color indicates the activity of each unit from 1 to 8.\\
Observe that the sequence and the pattern durations are shorter in the system with fast synapses ($\tau_r=300$) than the one with slow synapses ($\tau_r=900$). In the case of a weaker noise (Figure \ref{fig:parameterEffects}a) with fast synapses the system follows the sequence ABCDEF when $\mu=0.40714$ whereas it stops at the pattern B when $\mu=0.20714$. In other words, increasing $\mu$ in the system with fast synapses tends to recruits neurones sequentially. The system with slow synapses can follow the sequence ABCDEF in both cases. In fact the two different values of $\mu$ in each row correspond to the two different dynamical scenarios which have been evoked in the beginning of this section. This point will be developed in Section \ref{sec-dyn}. When noise is stronger (Figure \ref{fig:parameterEffects}b) the picture is different: the full sequence can be completed in the case of fast synapses even with $\mu=0.20714$, however in the case of slow synapses the sequence is short and the system quickly explores unexpected patterns like one with three excited units $4, 5, 6$ around $t=1000$ (which is not a learned pattern) in panel three. \\
Comparison between Figures \ref{fig:parameterEffects}b and \ref{fig:parameterEffects}c examplify the effect of changing $\lambda$ and $\mu$ for the same noise amplitude. Increasing the inhibition coefficient $\lambda$ regulates the transition for slow synapses, however fast synapses and high values of $\mu$ randomly activates learned patterns and yields short sequences (when however $\mu$ is smaller regular sequences can be preserved).

 \subsubsection{Length of a chain}
\begin{figure}
\centering
\includegraphics[scale=0.7]{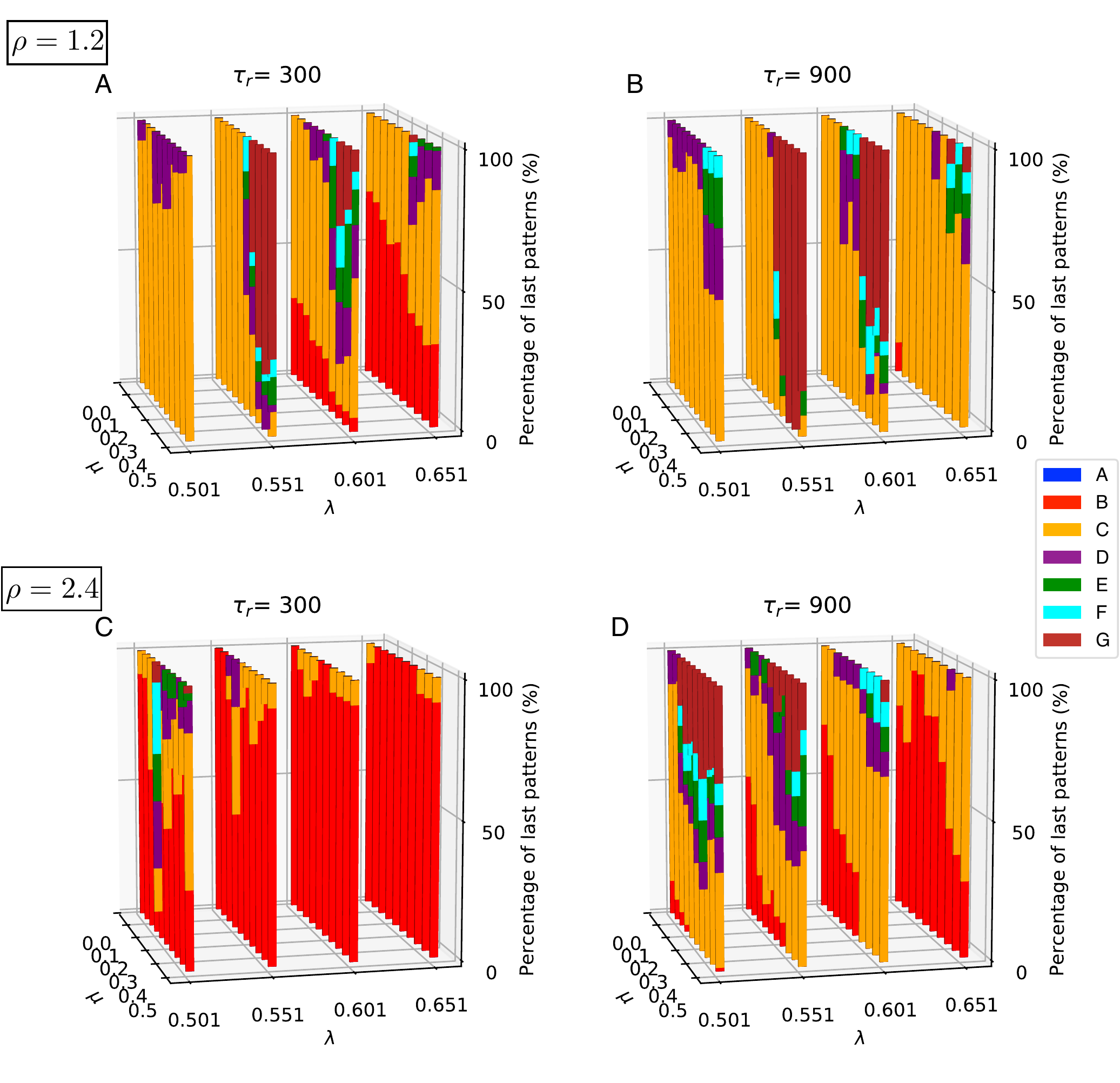}
\caption{Percentage of last activated patterns in a regular segment over simulations for $\eta=0.02$. Pattern colours follow to the colour codes of the last activated units (see the legend on the right). The height of each colour on a bar indicates the percentage of corresponding last activated pattern. For $\rho =1.2$ (panels A and B), the chain length increases with $(\mu, \tau_r)$. The global inhibition, $\lambda$, should be high enough for a sequential activation, but the chain length decreases if the inhibition is too strong. For $\rho =2.4$ (panels C and D), the chain length increases with $(\mu,\tau_r)$, but decreases with $\lambda$. The sharp increase in the chain length (more visibly in $\rho=1.2$) occurs when the bifurcation scenario changes around $\mu \approx \mu^*$.}
\label{fig:pattern_N=8_eta=20}
\end{figure}

\begin{figure}
\centering
\includegraphics[scale=0.7]{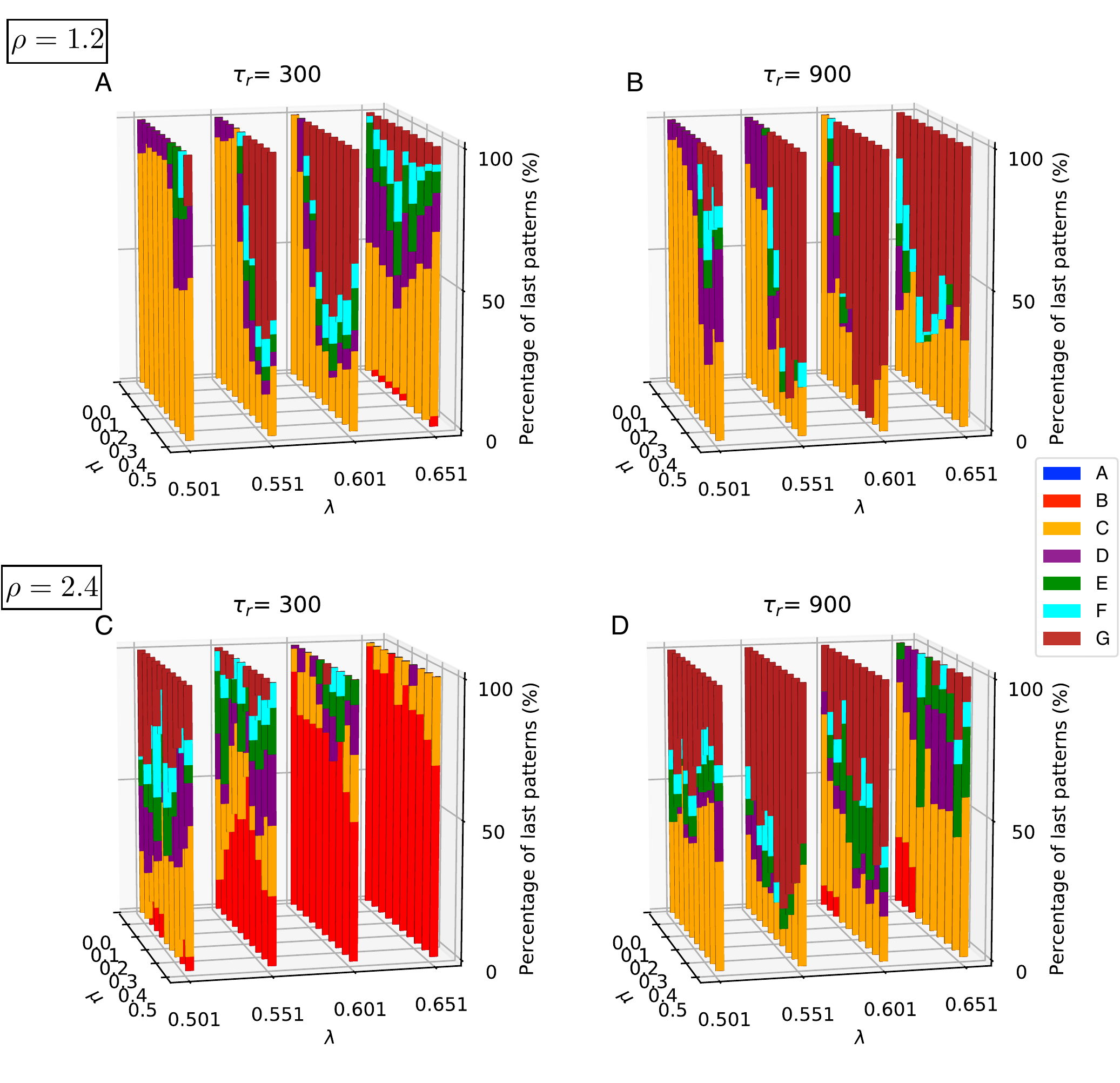}
\caption{Percentage of last activated patterns in a regular segment over simulations for $\eta=0.04$. Pattern colours follow to the colour codes of the last activated units (see the legend on the right). The height of each colour on a bar indicates the percentage of corresponding last activated pattern. For $\rho =1.2$ (panels A and B), the chain length increases with $\tau_r$. Chains are longer for intermediate values of $\mu$, but get shorter as $\mu$ increases. The global inhibition, $\lambda$, should be high enough for the sequential activation. Activation spreads over lower values of $\mu$ as $\lambda$ increases but the chain length can decrease if the inhibition is too strong. For $\rho =2.4$ (panels C and D), the chain length increases with $\tau_r$, but decreases with $\lambda$. Increasing $\mu$ prolongs the chains more with $\tau_r=900$ than  $\tau_r=300$. Activation is easier with $\rho =2.4$ than $\rho=1.2$ if $\tau_r =900$ and $(\mu, \lambda)$ are small, but the chains under strong inhibition are longer in $\rho = 1.2$ than $\rho=2.4$.}
\label{fig:pattern_N=8_eta=40}
\end{figure}

 When the patterns in a chain are explored in the right order by the system we call it {\em regular}. As we saw in Section \ref{sec:simulations} it can happen that only part of the full regular chain has been realised before it stops or starts exploring patterns in a different order, hence activating an irregular chain. We call the partial regular chain a {\em regular segment} and its length is the number of patterns it contains. 
Here we investigate the maximal length that a regular segment starting at pattern A can attain. This length is the rank of the last activated pattern over simulations. It depends on noise but also on the neuronal and synaptic parameters. In Figures \ref{fig:pattern_N=8_eta=20} and  \ref{fig:pattern_N=8_eta=40} we present statistics of this rank for two different noise intensities $\eta=0.02$ and $0.04$. Last activated patterns are represented by color bars, the length of which indicates the percentage of corresponding last activated pattern over 80 trials.
In each figure we took $\rho \in \{1.2, 2.4\}$,$\tau_r \in \{300, 900\}$. Parameters $\lambda$ and $\mu$ are varied wihin a range assuring the existence of chains of at least length 2.\\
Generally speaking, increasing the noise level facilitates the activation, hence, prolongs the chains. Especially for $\tau_r=900$ the chain length is considerably higher with $\eta=0.04$. We can also observe the difference between $\rho=1.2$ and $\rho=2.4$, for the former strong inhibition is more favourable but week inhibition is more favorable for the later .\\ 
Also note that when $\mu$ is small, an increase of $\mu$ provokes an increase of the length of the chain.
However, in most cases we find that the chain lengths are maximal for intermediate values of $\mu$.
This is clear intuitively: large gain (small $\mu$) prevents the units from deactivating, making the transition from one pattern to the next difficult. Small gain, on the other hand, prevents the next unit from activating. Another factor is the occurrence of the transition from scenario 1 to 2 (see Section \ref{sec-dyn} and Appendix \ref{app:2dplane}). 
The noise level $\rho$ and the global inhibition parameter $\lambda$ also influence the system's behaviour. For $\rho=1.2$, inhibition in the middle range leads to longer sequences, whereas weak inhibition is more suitable for $\rho=2.4$.


\subsection{Analysis of the dynamics}\label{sec-dyn}
\begin{figure}[ht]
\centering
\includegraphics[scale=0.22]{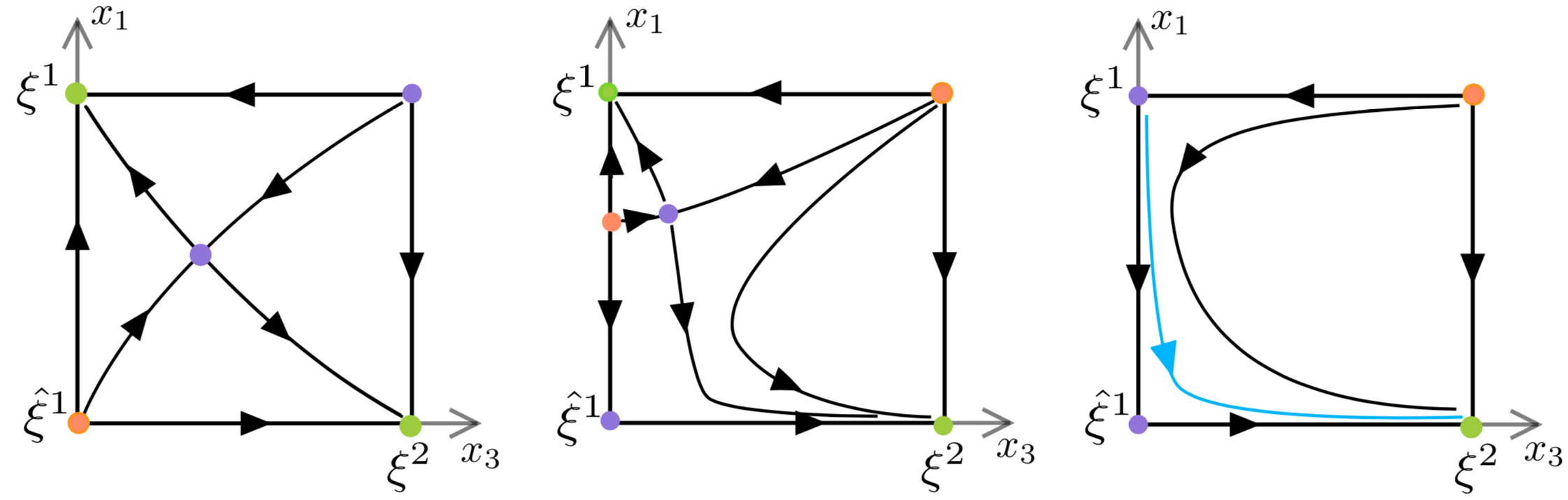}
\caption{Phase portrait of the fast dynamics on the face $F^2$ at three different 'slow' STD times $t<T_0$ (the learned patterns $\xi^1$ and $\xi^2$ are stable), $T_0<t<T_1$ (bifurcation of a saddle point on the edge between $\xi^1$ and $\hat\xi^1$) and $t>T_1$ ($\xi^1$ has become unstable along the edge $\xi^1-\hat\xi^1$ whilst $\xi^2$ is still stable). Green spots are the stable equilibriums, orange spots are completely unstable and purple spots are saddles. The blue lines illustrate segments of trajectory starting near  $\xi^1$. The saddle point in the interior of $F^2$ merges with the bifurcated equilibrium before right panel is realised.}
\label{fig:face}
\end{figure}

Latching dynamics is defined as a sequence (chain) of activations of learned patterns that de-activate due to a slow process (e.g., adaptation, here synaptic depression), allowing for a transition to the next learned pattern in the sequence (\cite{T05}, \cite{Lerner2012}).\\
Here we refine this description using the language of dynamics and multiple timescale analysis. The main idea is to treat the synaptic variables $s_i$ as {\em slowly varying parameters}, so that the evolution of the system becomes a \textit{movie} of the dynamical
configurations of the units $x_i$. On the other hand the firing rate equation \eqref{eq:xi'} is well adapted to analyze latching dynamics. Indeed from the form of \eqref{eq:xi'} (assuming for the moment that noise is set to 0) one can immediately see that whenever $x_i$ is set to $0$ or $1$, this variable stays fixed at any time. Therefore considering any face in the hypercube $[0,1]^N$ defined by two coordinates $x_i$, $x_j$, the other coordinates being fixed at $0$ or $1$, it is invariant under the flow of \eqref{eq:xi'}. In other words any trajectory starting in $F$ stays entirely in it. This is of course true also for the edges and vertices at the boundary of each face. Each vertex is an equilibrium of  \eqref{eq:xi'} and connections between such equilibria can be realised through edges of the hypercube, which greatly simplifies the analysis.\\ 
When the couple $(x_i, s_i)$ of unit $i$ is set at $(1,1)$, $x_i$ is fixed as we have seen but STD equation \eqref{eq:si'} induces an asymptotic decrease of the synaptic variable towards the value $S=(1+U\tau_r)^{-1}$. This in turn weakens the synaptic weight $J^{max}s_i$ in \ref{eq:xi'}, which may destabilize $\xi^i$ in the direction of $\hat\xi^i$. Considering $s_i$ as a slowly varying parameter this can be seen as a {\em dynamic bifurcation} of an equilibrium along the edge from $\xi^i$ to $\hat\xi^i$.
The following scenario was described in \cite{Aguilaretal2017}. For the sake of simplicity we now assume $i=1$ (the same arguments hold for any $i$). The patterns $\xi^1$, $\hat\xi^1$ and $\xi^2$ lie at the vertices of a face, which we call $F^2$, generated by the coordinates $x_1$ and $x_3$, with $x_2=1$ and the rest of the coordinates being set to 0.\\
Figure \ref{fig:face} shows three successive snapshots of the movie on $F^2$.
The left panel illustrates the initial configuration, with the stable pattern $\xi^1$ corresponding to the top left vertex. Then at some time $T_0$ an equilibrium bifurcates out of $\hat\xi^1$ in the direction of $\xi^1$ (here the 'slow' STD time plays the role of bifurcation parameter, see middle panel). After a time $T_1$ (right panel) this bifurcated equilibrium disappears in $\xi^1$ which becomes unstable and a connecting trajectory is created along the edge with $\hat\xi^1$. Simultaneously a trajectory connects $\hat\xi_1$ to $\xi_2$ on the corresponding edge. It results that the following sequence of connecting trajectories is created: $\xi^1\rightarrow\hat\xi^1\rightarrow\xi^2$. As a result, any state of the system initially close to $\xi^1$ will follow the 'vertical' edge towards $\hat\xi^1$, then the 'horizontal' edge towards $\xi_2$. The process can repeat itself from $\xi^2$ to $\xi^3$ and so on. It was shown in \cite{Aguilaretal2017} that in order to work, this scenario requires that the coefficients of the matrix $J^{max}$ satisfy the relation $J^{max}_{1,2}<J^{max}_{2,3}$ (more generally $J^{max}_{i,i+1}<J^{max}_{i+1,i+2}$, $i=1,\dots,P-1$, for the existence of a chain of $P$ patterns), a condition which does not hold with \eqref{Jmax}.

\begin{figure}[h!]
\centering
\includegraphics[scale=0.3]{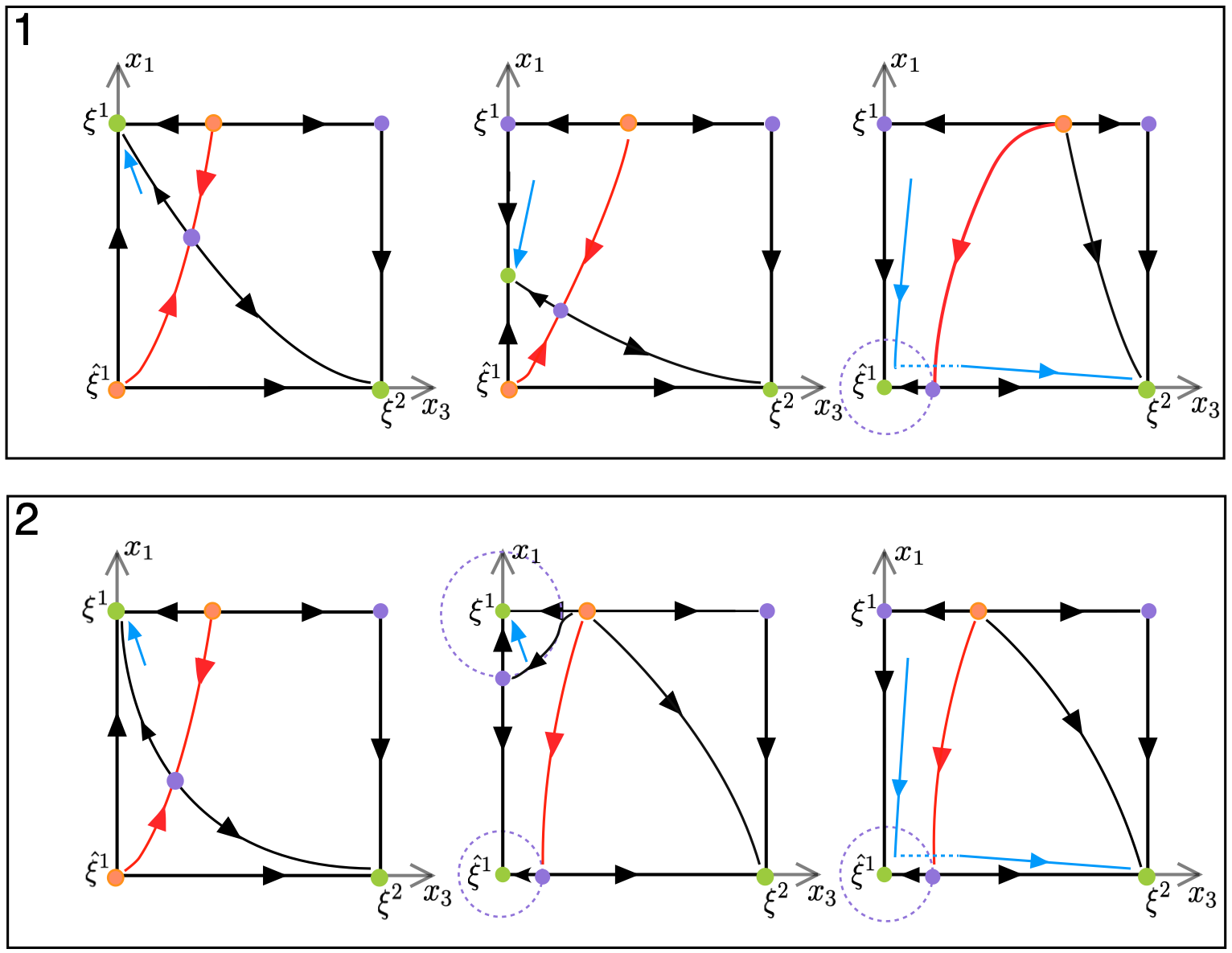}
\caption{Phase portraits of the fast dynamics on the face $F^2$ at three different 'slow' STD times, illustrating the transition $\xi^1\rightarrow\xi^2$ with excitable connections in Scenarios 1 (above) and 2 (below). Stable patterns are coloured in green. The red trajectories are separatrices between the basins of attraction of the stable equilibria. The blue lines illustrate segments of a trajectory starting near $\xi^1$. In panels 1 (c) and 2 (c) this trajectory "jumps" out of the basin of attraction of $\hat\xi^1$ under the effect of noise and converges towards $\xi^2$}
\label{fig:bifurcationODG}
\end{figure}

To circumvent this difficulty we relax the condition that the chain of connections $\xi^1\rightarrow\hat\xi^1\rightarrow\xi^2$ exists when $t>T_1$ (right panel of Fig. \ref{fig:face}). We assume instead that the connecting trajectory along the edge $\hat\xi^1-\xi^2$ is broken by a stable equilibrium close to $\hat\xi^1$. In such case strong enough noise perturbations could push a state of the system which has converged to $\hat\xi^1$ off its basin of attraction. As a result the system would escape $\hat\xi^1$ and converge towards $\xi^2$ (in a stochastic sense), as expected. When such chains driven by noise exist, we call them {\em excitable chains} by reference to \cite{ashwin-postlethwaite} who introduced the concept.\\
Under the new scheme the number of possible transitions is much larger and multiple outcomes are possible.
We have identified two scenarios (named 1 and 2) by which these excitable chains can occur in our problem. Typical cases are illustrated on Figure \ref{fig:bifurcationODG}. As in the previous figure, snapshots of the dynamics at three different "slow" times are shown.
The red line marks the boundary of the basin of attraction of $\xi^{2}$ and the dashed circles mark the closest distances for a possible stochastic jump out of it. In both scenarios a completely unstable equilibrium point exists on the edge from $\xi^1$ to the unnamed vertex on $F^2$, which corresponds to the pattern $(1,1,1,0,\dots,0)$ (not a learned pattern).\\ 
Under Scenario 1 the pattern $\xi^1$ loses first stability by a dynamic bifurcation of a sink (stable equilibrium) along the edge $\xi^1-\hat\xi^1$. This sink travels along the edge until it merges with $\hat\xi^1$, so that a heteroclinic connection from $\xi^1$ to $\hat\xi^1$ is realised. However a saddle equilibrium bifurcates from $\hat\xi^1$ along the edge $\hat\xi^1-\xi^2$ (middle panel). As a result $\hat\xi^1$ is weakly stable in the direction of $\xi^2$ and if noise is not too small the dynamics can jump to the basin of attraction of $\xi^2$ (right panel).\\ 
In Scenario 2 the picture is initially similar to that of Scenario 1. The difference comes from the simultaneous bifurcation of two equilibria from $\hat\xi^1$: one along the edge $\hat\xi^1-\xi^1$ and the other along the edge $\hat\xi^1-\xi^2$ (middle panel). The first bifurcated equilibrium eventually merges with $\xi^1$ so that a heteroclinic connection is created from $\xi^1$ to $\hat\xi^1$ by the same mechanism as in Figure \ref{fig:face} (right panel). However the other bifurcated equilibrium point prevents the formation of a heteroclinic connection from $\hat\xi^1$ to $\xi^2$. Nevertheless $\hat\xi^1$ is only weakly stable in the $\xi^2$ direction and noise can allow an incoming trajectory to jump over $\hat\xi^1$ towards $\xi^2$. 
Note that the equilibrium on the edge $\hat\xi^1-\xi^2$ can travel towards $\xi^2$ as $s_1$ elapses, so that the basin of attraction of $\hat\xi^1$ becomes too large for noise to allow jumps toward $\xi^2$. In this case the system may get indefinitely stucked at $\hat\xi^1$. We refer to this behaviour as {\em pending}, see an example in Fig. \ref{fig:parameterEffects}, second simulation in panel (a).\\
It is shown in Appendix \ref{app:dynamicBifurcation} that there indeed exist parameter domains in which one or the other of the two scenarios occurs.\\ 
Simulations (and analysis, see \ref{app:2dplane}) identify $\mu$ as the main control parameter which determines the choice between these scenarios: the system follows Scenario 1 for higher values of $\mu$ and Scenario 2 for lower values of $\mu$. This explains the difference in behavior seen in Figure \ref{fig:parameterEffects} at lower and higher values of $\mu$. 
The boundary between the two regions is defined by the value $\mu=\mu^*$ for which $\xi^i$ and $\hat\xi^i$ change stability at the same time.
For an analytic definition of $\mu^*$ and more detailed analysis, see \ref{app:2dplane}.

 \subsection{Irregular chains}\label{sec-more}
 
The question we address here is what happens after the last pattern of a regular segment has been reached. Here we discard the case when this last pattern remains stable indefinitely or when the system deactivates completely (all units relax to inactive state).  We focus on the case when the dynamics continues afterwards by jumping to patterns in the sequence in an irregular manner and generate new chains, possibly going backward.  
 Suppose at a time $t$, $x_p$ and $x_q$ are the two most recently activated units, with $x_p$ preceding $x_q$ in its activation. We define
  \[
 \Delta=q-p.
 \]
 Note that a regular chain satisfies $\Delta=1$ for all $t$ until the last pattern is reached.\\
 We distinguished two cases of irregular continuation of chains: reversing the chain ($\Delta=-1$) and random reactivation of new chains ($|\Delta|>1$). 
 Recall the scenarios 1 and 2 for transitions from one pattern to the next (fig. \ref{fig:bifurcationODG}). The former occurs for ''large'' values of $\mu$ and the latter for lower values of $\mu$. Let $\hat\xi_p$ be the last intermediate state at the end of the regular segment. In Scenario 2  either $\hat\xi_p$ remains stable indefinitely or it destabilizes after some (long) time due to the repotentiation of $s_p$. The latter case corresponds to a dynamic scenario for chain's reversal. We refer to the prolonged residence of the system at $\hat\xi_p$ as \textit{pending} and note that it likely leads to a reversal. However, for large values of noise, random activation ($|\Delta|>1$) may also occur. The scenario 1 is more likely to yield random re-activation as $\hat\xi_p$ loses stability in the $x^{p+1}$ direction with the decrease of $s_{p+1}$, so that a transition to the inactive state is possible. However in this case too other $\Delta$ values are possible when the noise is large.
  
\begin{figure}
\centering
\includegraphics[scale=0.27]{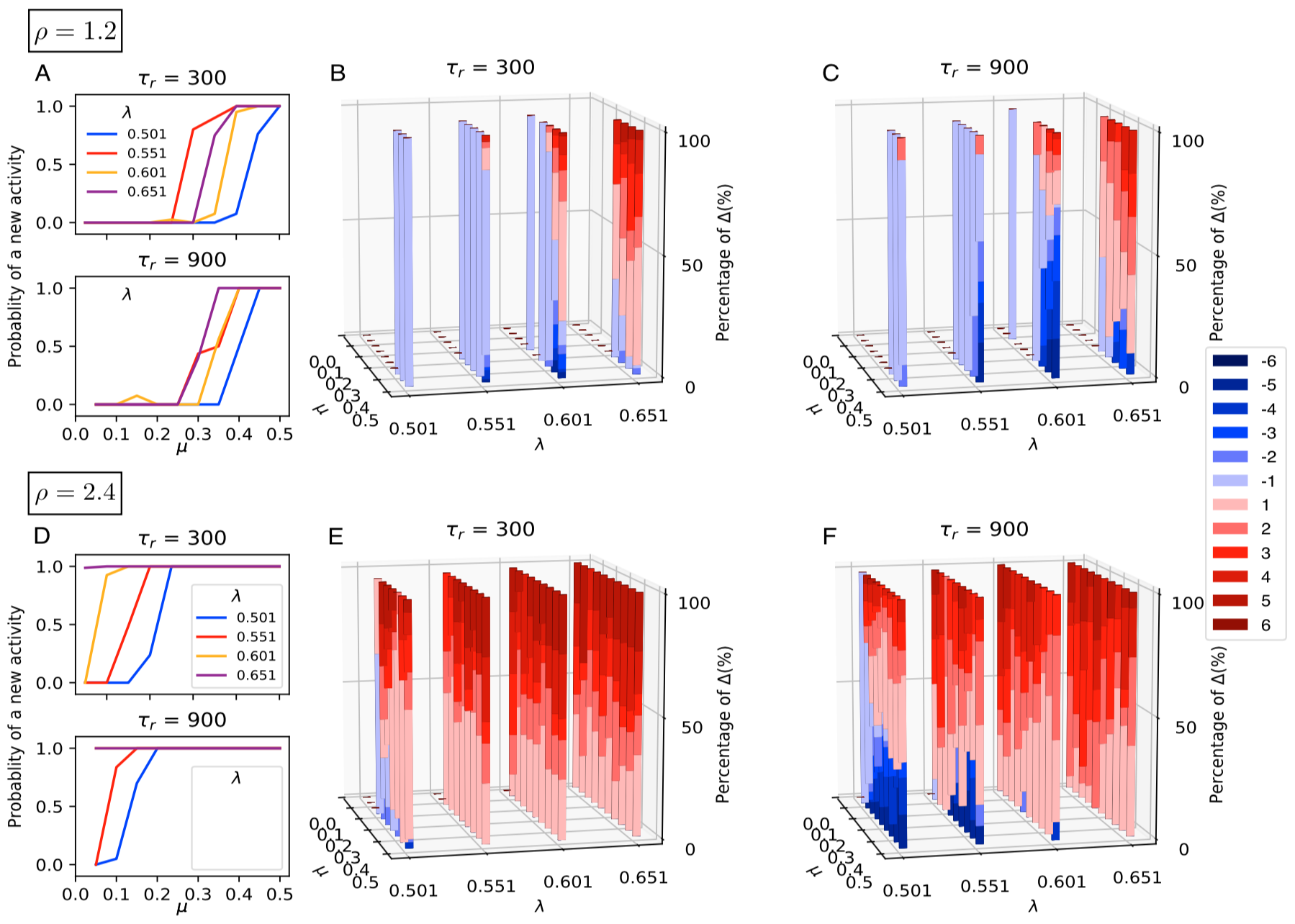}
\caption{Activity after the initial sequence for $\eta=0.02$. Panels A and D show the probability of a new activity to be observed after the initial sequence for $\rho=1.2$ and $\rho = 2.4$, respectively. Bar plots in panels B, C, E, and F show the percentage of activation distance $\Delta$ if there is any activity. Bars are coloured according to the distance colours (shown on the right) and the height of each colour indicates the percentage of the corresponding distance.
For $\rho=1.2$ (panels A, B and C) and small values of $\mu$ the system tends to remain on the last activated pattern, as low rates of activity probability in panel A demonstrate. New sequences are generated as $\mu$ and $\lambda$ increase, with a preference of backward activity for small values of $\lambda$. The probability of generating new sequences for small values of $\mu$ is higher if $\rho=2.4$ and the minimum value of $\mu$ required for a new sequence decreases with $\lambda$ and $\tau_r$ (panel D). For instance, if $\tau_r=900$ and $\lambda = \{0.601, 0.651\}$, new activity occurs in all trials. The new activity starts with distance$\Delta>0$ (panels E and F ) and we observe very few cases with $\Delta<0$ for $\rho =2.4$ (yet, more if $\tau_r=900$). The difference between the percentages of $\Delta$ for $\tau_r = 300$ and $\tau_r=900$ indicates the capability of slow synapses to yield longer chains.}
\label{fig:nextDistance_N=8_eta=20}
\end{figure}

\begin{figure}
\centering
\includegraphics[scale=0.27]{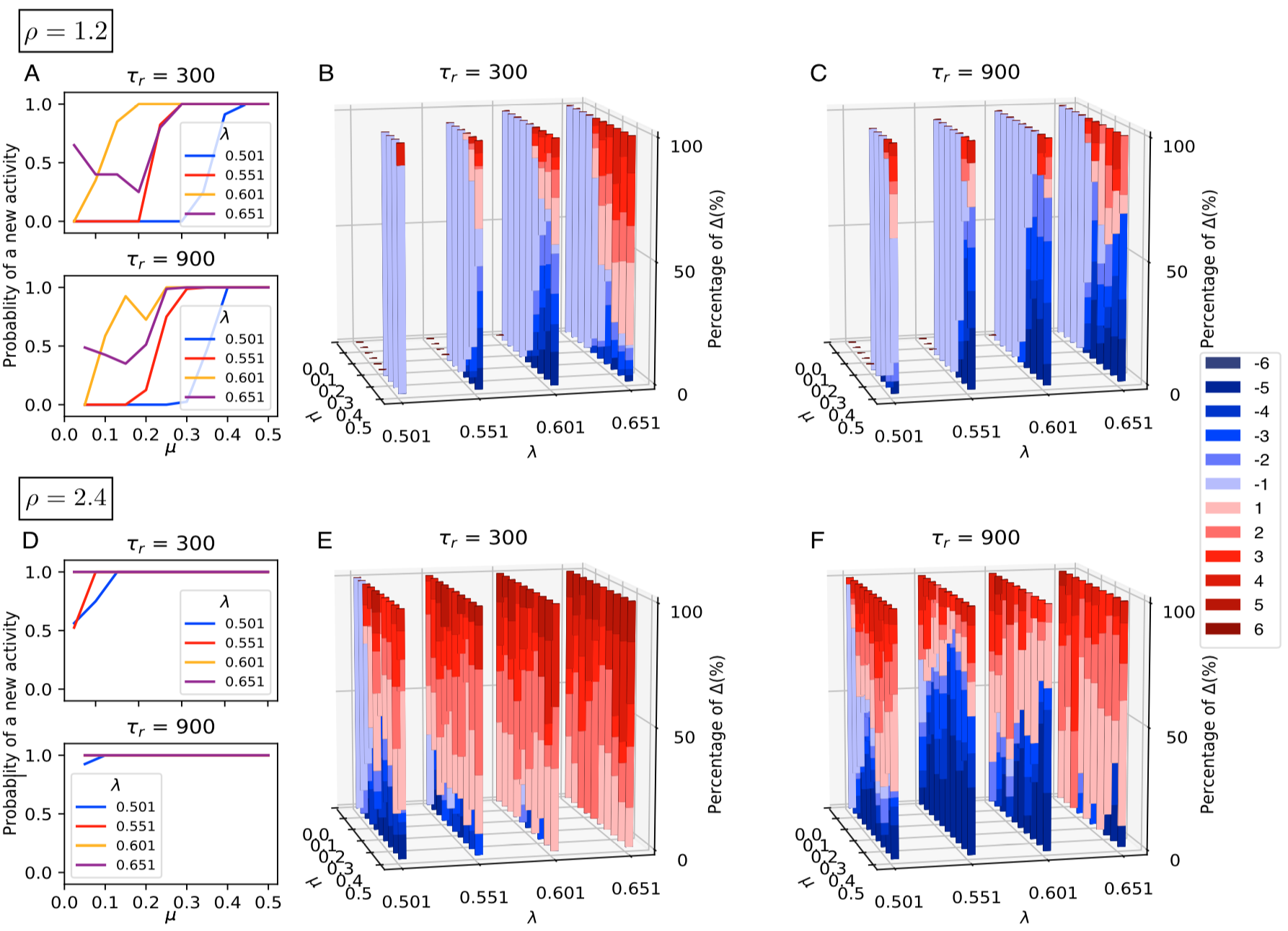}
\caption{Activity after the initial sequence for $\eta=0.04$. Same setting as in Fig. \ref{fig:nextDistance_N=8_eta=20}.
For $\rho=1.2$ (panels A, B and C), $\lambda = \{0.501, 0.551\}$ and small values of $\mu$, the system remains on the last activated pattern as low rates of activity probability in panel A demonstrate. New sequences are generated as $\mu$ and $\lambda$ increase, with a preference for backward activity for small values of $\lambda$. With $\rho=2.4$ (panels D, E and F) new activation can occur for all values of $\mu$ and $\lambda$. In panels E and F, increasing $\lambda$ and $\mu$ leads to an activation with $\Delta>0$. Percentage of activity with $\Delta<0$ is much higher with $\tau_r=900$ than $\tau_r=300$. The difference indicates the capability of slow synapses to yield longer chains both for $\rho=1.2$ and $\rho =2.4$.}
\label{fig:nextDistance_N=8_eta=40}
\end{figure}

Figures \ref{fig:nextDistance_N=8_eta=20} and \ref{fig:nextDistance_N=8_eta=40} show the percentage of $\Delta$ values after a new activation for $\eta=0.02$ and $\eta=0.04$, respectively. Activity with $\Delta=-1$ is generally supported for $\rho=1.2$ and if $\rho=2.4$, for small values of $(\mu, \lambda)$. As the chains get longer with increasing $\mu$, regular segments get longer, specially when noise is high ($\eta=0.04$). Regarding the type of forward and/or backward irregular chains, high values of $\rho$ and high values of $\mu$ (e.g. low gain) for low values of $\eta$ (Fig. \ref{fig:nextDistance_N=8_eta=20}) increase the possibility for irregular chains in the forward direction, while the combination of high values of $\mu$ (e.g. low gain), $\rho$ and $\eta$ increase the possibility for irregular chains in both directions (\ref{fig:nextDistance_N=8_eta=40}).

%
%
\newpage
\section{Discussion}

Experimental evidence indicates that the brain can either replay the same learned sequence to repeat reliable behaviors \cite{Velizcuba2015}, \cite{Conway2001}, \cite{Buhusi2005}, \cite{Eagleman2012}, \cite{Xu2012} or generate new sequences to create new behaviors \cite{Buckner2008}, \cite{Christoff2009}, \cite{Guilford1950}, \cite{Abraham2012}, \cite{Gonen2013} \cite{Fink2012}. The present research identifies biologically plausible mechanisms that explain how a neural network can switch from repeting learned regular sequences to activating new irregular sequences. To make the problem analytically tractable, the combined effects of the parameters were analyzed on neuronal population firing rates in a simplified balanced network model by use of slow-fast dynamics and dynamic bifurcations. We demonstrated how variations in neuronal gain, short-term synaptic depression and noise can switch the network behavior between regular or irregular sequences for a fixed learned synaptic matrix.\\

\subsection{Synaptic matrix}
In the present model the overlap had the same number of shared units for all the overlapping populations. This allowed to show that variable overlap is not a necessary condition for the activation of sequences of populations. A consequence of the constant overlap is that sequences from a stimulus-driven end-point pattern in the sequence (e. g. first pattern A of the sequence) are directional but sequences from a mid-point pattern can go in any of the two possible directions. The model can then generate bi-directional sequences interesting in free recall. Starting from the first pattern A (or G), the sequence ABCDEFG is oriented in one direction (or in the other direction), and starting from the first pattern e.g. D, the sequence can be oriented in any of the two possible directions.  
The present model allows for bi-directionnal sequences as well as for new sequences depending on the value of neuronal gain $\gamma=\mu^{-1}$.\\

\subsection{Regular $vs.$ irregular sequences}
Regarding regular sequences, the length of the chain activated increases with noise and for combinations of strong STD (high values of $\rho$) and low inhibition, or weak STD (low values of $\rho$) and strong inhibition. Further, for most combinations of noise, STD and inhibition, there is an optimal value of gain that generates the longest chains. The sensitivity of a neuron to its incoming activation varies with changes in its gain \cite{SalinasSejnowski2001}. Simulations and analysis show that the neuronal gain is a key control parameter that selects the length and type of sequence activated: regular or iregular. Large neuronal gain impairs the deactivation of the units in a pattern and hence makes the transition to the next pattern difficult, and small gain impairs the activation of the next unit and again makes the transition difficult. Consequently there is an optimal window for the gain corresponding to long sequences. Experimental evidence shows that presentations of a given stimulus reproduces the same sequence reliably \cite{Eagleman2012} \cite {Xu2012}; \cite{Shuler2006} \cite{Gavornik2014}. The present model can repeat systematic full sequences of activation for some values of the parameters that make the network change patterns in a given order. This 'reliable' mode could be well adapted to the reliable reproduction of learned sequences of behaviors.\\

Regarding irregular sequences, a large neuronal gain leads to the second scenario describing transitions from one pattern to the next (as in fig. \ref{fig:bifurcationODG}). According to this second scenario, the last intermediate state of the network at the end of a regular segment can destabilize and leads to a reversal of the sequence. In that case the network activates patterns backward in the reverse order. Further, for large values of noise, scenario two can lead to random activation of patterns in either the forward or backward direction. Such variable sequences are more likely to be generated according to the first scenario that makes possible a transition to another state in the forward or backward direction and that does not necessarily overlap with the current state (forward or backward leaps). Our model can generate variable sequences over repetitions of a same triggering stimulus for high values of gain, in line with a \emph{memoryless} system \cite{ashwin-postlethwaite} that activates a new pattern in an unpredictable fashion. Behavioral studies indicate that presentation of a triggering stimulus can activate distant items that are not directly associated to it \cite{Bowden2003}. The generation of new sequences corresponding to the activation of new possibilities \cite{Hassabis2007} and the execution of new information-seeking behaviors such as saccades or locomotor explorations of unknown locations \cite{Pezzulo2014}. This 'creative' mode of variable activation not following a given sequence could correspond to a mind wandering mode \cite{Christoff2009} \cite{AndrewsHanna2012} or divergent thinking involved in creativity \cite{Beaty2016} \cite{Guilford1967} \cite{Benedek2012} \cite{Benedek2013}.\\ 

\subsection{Neuromodulation of the switch between regular and irregular sequences}
A novel feature of our network model is that neuronal gain influences the type of sequences that are generated: regular or irregular. Typical computational models of sequence generation reproduce learned sequences \cite{Velizcuba2015}. However, if the brain must in some case reproduce systematic behaviors, it must also have the capacity to liberate itself from repetition in order to create new behaviors. The present research shows that the network can exhibit the dual behavior of activating regular or irregular sequences for a given synaptic matrix. The transition depends on biological parameters, in particular on gain modulation. Given that changes in gain change the length of the regular sequence, and that when the regular sequence stops it becomes irregular, the gain controls the regularity of the sequences. The gain is reported to depend on neuromodulatory factors such as dopamine \cite{Rolls2008} \cite{Seamans2001}\cite{Braver1999} \cite{Cohen1992} involved in reward-seeking behaviors and punishment \cite{Jhou2019} \cite{Lak2014} \cite{Pessiglione2006}. Dopamine is reported to modulate the magnitude of the activation between associates in memory (priming; \cite{Roesch2006} \cite{Kischka1996}) and dopamine induced changes in neuronal gain have been reported to account for changes in activation in memory \cite{Lavigne2008} and for movement control \cite{Stroud2018}. The present research sheds light on how the brain can switch between a 'reliable' mode and a 'creative' mode of sequencial behavior depending on external factors such as reward that neuromodulate neuronal gain.\\


\section*{Acknowledgements}
EKE and MK were supported by the ERC Advanced Grant NerVi no. 227747.\\[0.2ex]
The authors thank Gianluigi Mongillo for helpful discussions.


\appendix

\section{\label{app:dynamicBifurcation} Latching dynamics from the slow-fast view point}

 Throughout this work we have assumed that the firing rates $x_i$ evolve on a faster time scale than the 
 synaptic variables $s_i$. This can be formalized by redefining
 \eqref{eq:xi'} as a slow-fast system:
\begin{equation}\label{eq:slowfast}
\begin{split}
\dot{x_i} &= x_i(1-x_i)\left(-\mu x_i - I- \lambda \sum_{j=1}^N x_j+ \sum_{j=1}^N J_{i,j}^{max} s_j x_j \right)+\eta\\
\dot{s}_i &= \eps((1-s_i)-\rho s_i x_i)
\end{split}
\end{equation}
where $\displaystyle{\eps = 1/\tau_r}$ and $\rho = \tau_r U$ is the parameter introduced in Section \ref{sec:model}. 
To keep our presentation consistent with \cite{Aguilaretal2017} we use a slightly more general version of the model,
adding the parameter $I$ that can be understood as feedforward inhibition or as modulation of the excitability of the $i$th unit.
Using the formulation \eqref{eq:slowfast} we can apply the tools of the slow-fast systems' theory to gain insight into the transitions between learned patterns. Here we will make some basic observations, leaving rigorous slow-fast analysis for a later publication.
\subsection{\label{app:2dplane}Fast subsystem dynamics in the 2D transition plane}
By setting $\eps = 0$ in \eqref{eq:slowfast}, we obtain the fast subsystem of \eqref{eq:slowfast} where $s_i$'s are considered as parameters. 
This underlies the idea of dynamic bifurcation: as the $s_i$'s change, the features of the dynamics of the fast system,
in particular the stability properties of the patterns $\xi_i$,  evolve.
We keep in mind, however, that the $s_i$'s must follow the slow flow, so their values are not arbitrary and are related to each other.
We are particularly interested in transitions from $\xi_i$ to $\xi_{i+1}$, with the dynamics passing near $\hat\xi_i$.
During this transition $x_i$ moves from near $1$ to near $0$,  $x_{i+1}$ stays close to $1$, and $x_{i+2}$ moves from
near $0$ to near $1$. The remaining units stay near $0$.
Hence the relevant dynamics is approximated by the restriction of the fast system to the plane :
\[
F^{i+1}=\{(x_1,\ldots, x_N)\; :\; x_{i+1}=1, \; x_j=0\mbox{ if } j\neq \{i, i+2\}\},
\] 
given by:
\begin{equation}\label{eq-sf2D}
\begin{split}
\dot x_{i} & = x_{i}(1-x_{i})\left(-\mu x_{i}-I-\lambda(1+x_{i}+x_{i+2})+2s_{i}x_{i}+s_{i+1}\right)\\
\dot x_{i+2} & = x_{i+2}(1-x_{i+2})\left(-\mu x_{i+2}-I-\lambda(1+x_{i}+x_{i+2})+2s_{i+2}x_{i+2}+s_{i+1}\right).
\end{split}
\end{equation}
Since $0\le x_i\le 1$
we are interested in the dynamics of \eqref{eq-sf2D} restricted to the square $[0, 1]^2$, whose edges are invariant for the dynamics. The equilibrium points $(x_{i+2}, x_{i})=(0,1)$ and $(x_{i+2}, x_{i})=(1,0)$ represent $\xi^{i}$ and $\xi^{i+1}$, respectively, while $(x_{i+2}, x_{i})=(0,0)$ represents $\hat\xi^{i}$ (the relevant picture is given in Fig \ref{fig:bifurcationODG}, with the subscripts $1$ and $2$
replaced by $i$ and $i+1$). 
\subsection{Dynamic bifurcation scenarios}
As $s_i$ and  $s_{i+1}$ vary in $[S,1]$
(see Section \ref{sec:model} for the definition of $S$), 
the equilibria of \eqref{eq-sf2D} undergo several bifurcations that are responsible for the transition $\xi^i \to \hat\xi^i \to \xi^{i+1}$,
which, in the particular context of \eqref{eq-sf2D}, occurs when the eigenvalues of $\xi_i$ or $\hat\xi_i$ 
corresponding to the edge $x_{i+2}=0$ pass through $0$.
These eigenvalues can be easily read off from \eqref{eq-sf2D}:
\begin{equation}
\begin{split}
&\lambda_i=-(\mu +2 \lambda + I)+2s_i+s_{i+1}\\
&\hat\lambda_i=-(\lambda+I)-s_{i+1}.
\end{split}
\end{equation}
Note that both $\lambda_i$ and $\hat\lambda_i$ are time dependent and vary on the same time scale as the $s$'s.
According to the above we are interested in the following bifurcations: 
\begin{itemize}
\item
The initially stable equilibrium point $(x_{i+2}, x_{i})=(0,1)$ becomes unstable at $$2 s_{i}(t_{(0,1)}) + s_{i+1}(t_{(0,1)}) = \mu +2 \lambda + I,$$ where $t_{(0,1)}$ is the corresponding bifurcation moment.
\item Initially unstable equilibrium point $(x_{i+2}, x_{i})=(0,0)$ becomes stable at $$s_{i+1}(t_{(0,0)})= \lambda + I,$$ where $t_{(0,0)}$ is the corresponding bifurcation moment.
\end{itemize}
The order of these bifurcations depends on the values of $\mu, \lambda$ and $I$ and we identified two different 
dynamic bifurcation scenarios: Scenario 1 if $t_{(0,1)} < t_{(0,0)}$ and  Scenario 2 if $t_{(0,0)} < t_{(0,1)}$.

The  phase portraits defining Scenario 1 are sketched in panel 1 of Figure \ref{fig:bifurcationODG} and can be characterized as follows:
\begin{itemize}
\item Phase portrait (a) in panel 1 corresponds to the situation where $\lambda<0$ and $(0, 1)$ is stable. At the same 
time $\hat\lambda>0$ and $(0, 0)$ is unstable. 
\item Phase portrait (b) in panel 1 corresponds to the case  
$\lambda_i>0$ and $\hat\lambda_i>0$. 
In this case $(0, 1)$ has become unstable and there exists a stable equilibrium
on the $x_{i}$-axis, $(x_{i+2}, x_{i})=(0, x_i^*)$. The point $(0, 0)$ is still unstable.
\item  
$\lambda_i>0$ and $\hat\lambda_i<0$. In this case $(0, 0)$ becomes stable and there exists
a saddle at $(x_{i+2}^*,0)$. There is a possibility of an \textit{excitable} connection, when trajectories travel down to $(0, 0)$,
jump over to $(x_{i+2}^*,0)$ by the action of noise, and continue on to $(1, 0)$.
\end{itemize}
\begin{figure}
\center
\includegraphics[scale=1]{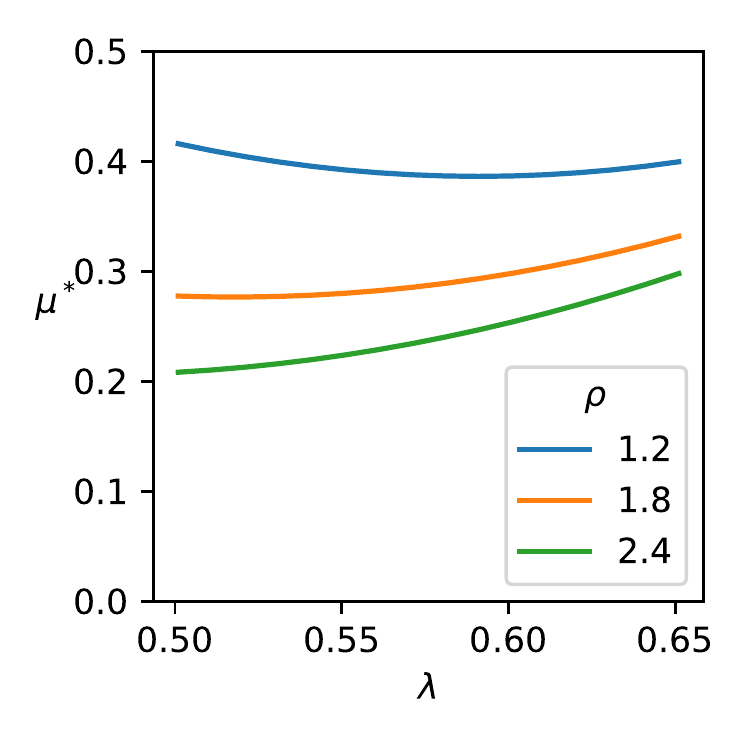}
\caption{Variation of $\mu^*$ versus $\lambda$ for $\rho=\{1.2, 1.8, 2.4\}$. The function $\mu^*(\lambda)$ changes non-monotonically for $\rho=\{1.2, 1.8\}$ with minima at $(\lambda, \mu)=(0.591, 0.3863)$ and $(\lambda, \mu)=(0.521, 0.2768)$, respectively, but increases monotonically for $\rho = 2.4$.}
\label{fig:muCritical}
\end{figure}
The  phase portraits defining Scenario 2 are sketched in panel 2 of Figure \ref{fig:bifurcationODG} and can be characterized as follows:
\begin{itemize}
\item Phase portrait (a) in panel 2 corresponds to the situation where $\lambda_i<0$ and $(0, 1)$ is stable. At the same 
time $\hat\lambda>0$ and $(0, 0)$ is unstable. 
\item Phase portrait (b) in panel 2 corresponds to the case  when both 
$\lambda_i$ and $\hat\lambda_i$ are negative. 
In this case $(0, 1)$ is still stable, while $(0, 0)$ has become also stable and there exists a saddle type equilibrium
on the $x_{i}$-axis, $(x_{i+2}, x_{i})=(0, x_i^*)$. 
\item 
Phase portrait (c) in panel 2 corresponds to the case  
$\lambda_i>0$ and $\hat\lambda_i<0$. In this case $(0, 1)$ has become unstable and 
trajectories can pass from $(1, 0)$ to $(0, 0)$ and jump over  a saddle point on the line $x_i=0$,
making an excitable connection to $(0, 1)$.
\end{itemize}
The feature that distinguishes Scenarios 1 and 2 is that the saddle $(0, x^*_{i+2})$ present in Scenario 1 bifurcates dynamically from 
$(0, 0)$. Hence the jump required is arbitrarily small. 

Near the bifurcation of $(0,0)$ at $\hat\lambda_i = 0$, $s_{i+1} = I+\lambda$ and $ s_{i+2}\approx 1 $, whereas the value of $s_{i}$ is harder to compute, but we know that
\[
S<s_{i}<s_{i+1} = I+\lambda.
\]
Note that when $\hat\lambda = 0$ (equivalently $s_{i+1} = \lambda +I$), the equilibrium at $(0, 0)$, corresponding to $\hat\xi^i$ for the original system, has two $0$ eigenvalues.

We now determine $\mu=\mu^*$ which separates Scenario 1  from Scenario 2. Note that $\mu^*$ is 
defined by the requirement $t_{(0,0)}=t_{(1,0)}$. We will count the time $t$ starting with the previous transition
$\hat\xi_{i-1}\to\xi_i$ and we will assume that this transition is instantaneous.
Note that at that time $s_{i+1}\approx 1$ (this holds for any $\mu$). Given that $s_{i+1}(t_{(0,0)}) = \lambda +I$ we obtain,
using the slow equation:
\[
t_{(0,0)}=\int_{1}^{\lambda+I} \frac{ds}{\eps(1-s(1+\rho))}.
\]
Using the assumptions $t_{(0,0)}=t_{(1,0)}$ and $\hat\lambda_i(t_{(1,0)})=0$ we obtain
$2s_i(t_{(0,0)})=\lambda+\mu$. Given that the computation of $s_i$ is independent of $i$ we now that $s_i=\lambda+I$ at the time of the transition
$\hat\xi_{i-1}\to\xi_i$. Hence
\[
t_{(0,0)}=t_{(1,0)}=\int_{\lambda+I}^{\frac{\mu+\lambda}{2}} \frac{ds}{\eps(1-s(1+\rho))}
\]
This implies that $\mu^*$ is given by the following equation:
\begin{equation}\label{eq:muboundary}
\int_{1}^{\lambda+I} \frac{ds}{\eps(1-s(1+\rho))} = \int_{\lambda+I}^{\frac{\mu^*+\lambda}{2}} \frac{ds}{\eps(1-s(1+\rho))}
\end{equation}
Decreasing $\mu$ from $\mu^*$ gives $t_{(1,0)}>t_{(0,0)}$, i.e. Scenario 2 and increasing $\mu$ from $\mu^*$ gives $t_{(0,0)}>t_{(1,0)}$,
i.e. Scenario 1.

There are some additional constraints that are shared between the two scenarios.
First, we require that  $\hat \xi^i$ should be stable in the transverse directions, in particular in the direction of $x_{i+1}$. The relevant eigenvalue is $\hat \sigma_{i+1}=\mu+I+\lambda-2s_{i+1}$, which gives, upon substitution of $s_{i+1}=I+\lambda$, $\hat \sigma_{i+1}=\mu-I-\lambda$. This introduces another condition:
\begin{equation}\label{b1cond2}
\mu <\lambda+I.
\end{equation}

Second, as in \cite{Aguilaretal2017}, we require the stability of $\xi^i$ in the absence of synaptic depression and the stability of $\hat\xi^{i}$ in transverse directions. This implies:
\begin{equation}\label{stacond}
I+2\lambda+\mu< 2,\quad I+\lambda<1<I+2\lambda.
\end{equation}

Figure \ref{fig:muCritical} shows the $\mu^*$ decreasing with $\rho$: $\mu^*$ increases with $\lambda$ for $\rho=2.4$, whereas its minimum is in the middle ranges of $\lambda$ for $\rho=1.2$. The form of the $\mu^*(\lambda)$ function explains why the system gives longer chains under weak inhibition for $\rho=2.4$, but middle/strong inhibition for $\rho=1.2$ in Figures \ref{fig:pattern_N=8_eta=20} and \ref{fig:pattern_N=8_eta=40}.

\end{document}